\documentclass[aps,prb,10pt,twocolumn,superscriptaddress,nofootinbib,preprintnumbers,showpacs]{revtex4-2}

\usepackage{amsmath,amsfonts, amssymb, amsthm, dsfont}
\usepackage[T1]{fontenc}
\usepackage{yfonts}
\usepackage{bm}
\usepackage{bbm}
\usepackage{mathrsfs}
\usepackage{graphicx}
\usepackage{verbatim}
\usepackage[bookmarks=true,colorlinks=true,linkcolor=blue, urlcolor=blue, citecolor=blue]{hyperref}
\hypersetup{linktocpage}
\usepackage{wasysym}
\usepackage[dvipsnames]{xcolor}
\usepackage{bbold}
\usepackage{braket}
\usepackage{diagbox} 
\usepackage{environ} 
\usepackage{relsize} 
\usepackage[caption=false]{subfig} 
\usepackage{tikz}
\usepackage{ulem}
\usepackage[greek,english]{babel}
\usepackage{teubner}

\usepackage[dvipsnames]{xcolor}

\newcommand{\qq}{\text{ϙ}}

\newcommand{\XY}[1]{\textit{{\textcolor{blue}{#1}}}}

\newcommand{\XW}[1]{\textcolor{NavyBlue}{#1}}

\begin{document}
\title{\mbox{Logarithmic corrections to bulk and surface criticality in a three-dimensional }
\mbox{quantum Heisenberg antiferromagnet}} 
\author{Xuyang Liang}
\affiliation{Guangdong Provincial Key Laboratory of Magnetoelectric Physics and Devices, State Key Laboratory of Optoelectronic Materials and Technologies, Institute of Neutron Science and Technology, School of Physics, Sun Yat-Sen University, Guangzhou, 510275, China}

\author{Xiao-Chuan Wu}
\email{xw0618@princeton.edu}
\affiliation{Department of Physics, Princeton University, Princeton, NJ 08544, USA}

\author{Zenan Liu}
\affiliation{Department of Physics, School of Science and Research Center for Industries of the Future, Westlake University, Hangzhou 310030,  China}
\affiliation{Institute of Natural Sciences, Westlake Institute for Advanced Study, Hangzhou 310024, China}

\author{Zhe Wang}
\affiliation{Department of Physics, School of Science and Research Center for Industries of the Future, Westlake University, Hangzhou 310030,  China}
\affiliation{Institute of Natural Sciences, Westlake Institute for Advanced Study, Hangzhou 310024, China}

\author{Zheng Yan}
\email{zhengyan@westlake.edu.cn}
\affiliation{Department of Physics, School of Science and Research Center for Industries of the Future, Westlake University, Hangzhou 310030, China}
\affiliation{Institute of Natural Sciences, Westlake Institute for Advanced Study, Hangzhou 310024, China}

\author{Dao-Xin Yao}
\email{yaodaox@mail.sysu.edu.cn}
\affiliation{Guangdong Provincial Key Laboratory of Magnetoelectric Physics and Devices, State Key Laboratory of Optoelectronic Materials and Technologies, Institute of Neutron Science and Technology, School of Physics, Sun Yat-Sen University, Guangzhou, 510275, China}

\begin{abstract}

At the bulk upper critical dimension, marginally irrelevant interactions generate multiplicative logarithmic corrections to mean-field scaling. While these corrections are well understood for bulk observables, their consequences for boundary criticality, particularly for finite-size scaling, remain much less explored. Here we combine large-scale quantum Monte Carlo simulations with boundary renormalization-group analysis to study a $(3+1)$D O(3) quantum critical point. 
After verifying the known logarithmically modified bulk finite-size scaling, including the correlation-length scaling governed by the logarithmic finite-size exponent $\hat{\qq}$, we tune the surface coupling to identify ordinary, special, and extraordinary boundary regimes. 
For the ordinary and special transitions, we derive logarithmic correction exponents and $\hat{\qq}$-dependent finite-size scaling forms for boundary correlations, including results that have not been systematically established before. These predictions are quantitatively supported by Monte Carlo data. In the extraordinary regime, we find long-range surface magnetic order and a logarithmically enhanced surface-bulk correlation.

\end{abstract}



\maketitle

\XY{Introduction.} A critical bulk can support distinct boundary universality classes depending on the boundary interactions. In the conventional theory of boundary critical phenomena (see Refs.~\cite{Binder1983Critical,Diehl1986Field,Diehl1997theory} and references therein), the ordinary transition corresponds to a bulk critical point with a disordered boundary, whereas the extraordinary transition corresponds to a bulk critical point with an ordered boundary. These two regimes are separated by the special transition, a boundary multicritical point. Boundary observables are characterized by their own critical exponents and scaling relations, reflecting the interplay between boundary and bulk critical fluctuations. Recent studies have shown that this interplay can lead to new boundary critical behavior in $(2+1)$D $\textrm{O}(N)$ models, such as the extraordinary-log phase~\cite{Metlitski2022boundary,Padayasi2022the,Parisen2021boundary,Parisen2022boundary,Hu2021extraordinary,Sun2023extraordinary,parisen2024universal}.

At the bulk upper critical dimension, this interplay acquires an additional subtlety. Marginally irrelevant interactions generate multiplicative logarithmic corrections to mean-field scaling. For bulk critical phenomena, these corrections and the associated logarithmic critical exponents are well understood~\cite{Wegner1973Logarithmic,zinn2021quantum,KENNA1993Renormalization,Kenna1994scaling,KENNA2004finite,Kenna2006scaling,Kenna2006Self-Consistent,Janssen2004Logarithmic,bauerschmidt2014scaling,aizenman1983renormalized,ruiz2017revisiting}. Finite-size scaling contains a further ingredient: the finite-size correlation length obeys
$\xi\sim L(\ln L)^{\hat{\qq}}$,
where $\hat{\qq}$ is the logarithmic finite-size exponent~\cite{aktekin2001finite,KENNA2004finite,Kenna2006scaling,Kenna2006Self-Consistent,Kenna2013Anew,Kenna2014Fisher}. This modified finite-size length scale plays an essential role in the scaling of bulk observables and has been extensively investigated in systems at the upper critical dimension~\cite{Fang2021Logarithmic,Li2024Logarithmic,Lv2021Finite}. By contrast, logarithmic corrections to boundary criticality, particularly their consequences for finite-size scaling, remain much less explored. A natural question is how the logarithmically slow running of marginally irrelevant bulk interactions affects boundary critical behavior, and how the resulting logarithmic corrections enter the finite-size scaling of boundary observables.

To address this issue, we study a spin-$1/2$ columnar-dimerized quantum Heisenberg antiferromagnet whose bulk transition realizes a $(3+1)$D $\mathrm{O}(3)$ quantum critical point~\cite{liang2025scaling,nohadani2005quantum,Oitmma2012Universal}. Using large-scale quantum Monte Carlo (QMC) simulations~\cite{sandvik1991quantum,sandvik1999stochastic,syljuaasen2002quantum,yan2019sweeping,yan2022global}, we first verify the bulk logarithmic scaling in this lattice model, including the finite-size correlation length controlled by $\hat{\text{ϙ}}$. We then vary the surface coupling to map out the boundary regimes and locate the special transition separating the ordinary and extraordinary sides. Guided by boundary renormalization-group (RG) analysis, we derive the logarithmic correction exponents for boundary correlations and show how the logarithmic length scale governed by $\hat{\text{ϙ}}$ enters boundary finite-size scaling. This yields $\hat{\eta}_{\parallel}$ and $\hat{\eta}_{\perp}$ at the ordinary and special transitions, together with $\hat{\text{ϙ}}$-dependent scaling forms. To our knowledge, the special-transition exponents and these finite-size formulas have not been reported previously. The predictions are borne out by the QMC data. On the extraordinary side, our simulations reveal long-range surface magnetic order and a logarithmically enhanced surface-bulk correlation.

\begin{figure}
    \centering
    \includegraphics[width=0.7\linewidth]{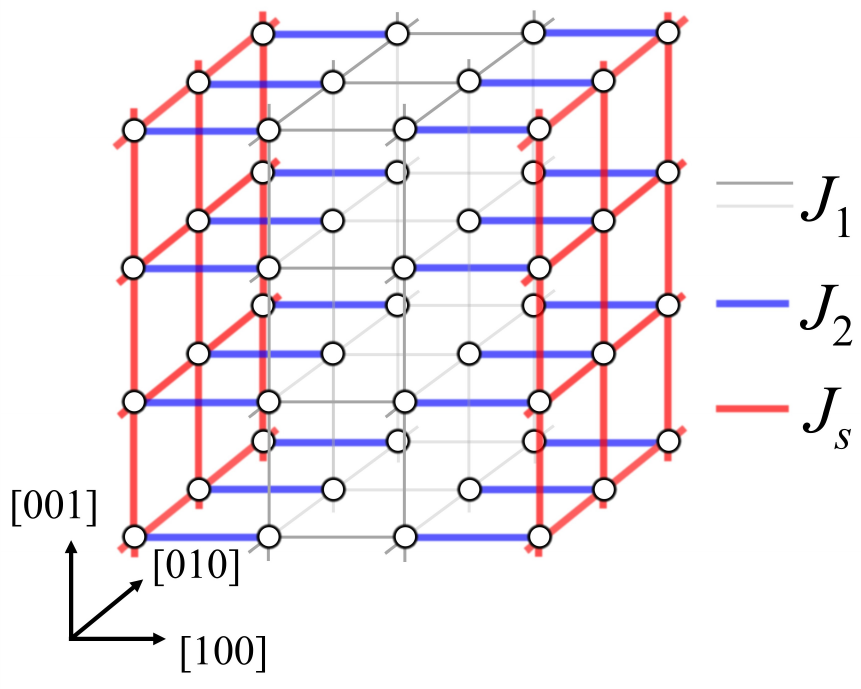}
    \caption{The columnar-dimerized lattice of the quantum Heisenberg antiferromagnet, where $J_1$ and $J_2$ denote the weak and strong bulk exchange couplings, respectively, and $J_s$ is the exchange coupling within the boundary layers.}
    \label{fig:model}
\end{figure}

\XY{Model and method.} We consider the spin-$1/2$ columnar-dimerized quantum Heisenberg antiferromagnet illustrated in Fig.~\ref{fig:model}. The Hamiltonian is written as
\begin{equation}
\begin{array}{l}
\begin{aligned}
H=\;& J_1\sum\limits_{\left \langle i,j\right \rangle}{{{S}}_{i} \cdot {{S}}_{j}}+J_2\sum\limits_{\left \langle i,j \right \rangle'}{{S}}_{i} \cdot {{S}}_{j}+J_s\sum\limits_{\left \langle i,j \right \rangle_s}{{S}}_{i} \cdot {{S}}_{j},
\end{aligned}
\end{array}
\label{Eq:Hmlt}
\end{equation} 
where $J_2$ and $J_1$ represent the strong and weak interactions in the bulk, with $J_s$ denoting the coupling strength of surfaces. In practice, we set $J_1= 1$ as the unit of energy. By tuning $J_2$, the model undergoes a quantum phase transition from an AFM N\'{e}el phase to disordered dimerized phase at $g_c$=4.0159(1) ($g=J_2/J_1$)~\cite{liang2025scaling,nohadani2005quantum,Oitmma2012Universal}. In this work, we employ stochastic series expansion (SSE) ~\cite{sandvik1991quantum,sandvik1999stochastic} QMC at inverse temperature $\beta=2L$ to investigate the bulk and boundary critical behavior. The order parameter staggered moment is defined as
$M_{q,\Omega}^z=\sum_{i\in \Omega }e^{-iq\cdot r_i}S_i^z$,
where $\Omega$ contains either all spins in the system or the spins on a surface, $r_i$ represents the real-space position of the spin $S_i$ at lattice site $i$, and $q$ is the wave vector. The corresponding bulk and surface staggered moments are denoted by $M_{q}^z$ and $M_{q,s}^z$. Based on the Kubo formula \cite{sandvik1991quantum,sandvik2010computational}, the bulk susceptibility is calculated from 
$\chi(q)=\frac{1}{L^3} \int_{0}^{\beta} d\tau \left \langle M_{-q}^z(\tau)M_{q}^z(0) \right \rangle$.
The correlation length $\xi$ is evaluated as
$\xi=\frac{L}{2\pi}\sqrt{\frac{\chi(Q)}{\chi(Q+\delta q)}-1}$,
where $\delta q=(2\pi/L,0,0)$ is one of the wave vectors closest to $q$ along the $x$ direction. The bulk and surface staggered magnetizations are defined, respectively, as
$m^z=\sqrt{\left \langle ({M_{Q}^z}/{L^3})^2\right \rangle }$ and $m_s^z=\sqrt{\left \langle ({M_{Q,s}^z}/{L^2})^2\right \rangle}$,
where $Q$ denotes the wave vector of AFM order. The surface Binder ratio is given by
$R_2={\left \langle (M_{Q,s}^z)^4 \right \rangle}/{\left \langle (M_{Q,s}^z)^2 \right \rangle^2}$.
We locate the special transition from the finite-size crossings of $R_2$, which is dimensionless at criticality. To further probe the universal properties of surface criticality, we measure correlation functions $C_{\parallel(\perp )}(r)=\left \langle S_{r}^{z}S_{0}^{z}\right \rangle$, where $S_0^z$ is taken as a reference spin on the surface.  The parallel correlation $C_{\parallel}(r)$ is measured between two surface spins separated by a distance $r$ along the surface, while the perpendicular correlation $C_{\perp}(r)$ is measured between a surface spin and a bulk spin displaced by a distance $r$ along the direction normal to the surface. These two quantities are analyzed within the finite-size scaling framework to determine the corresponding surface critical exponents.

\begin{figure}
    \centering
    \includegraphics[width=1\linewidth]{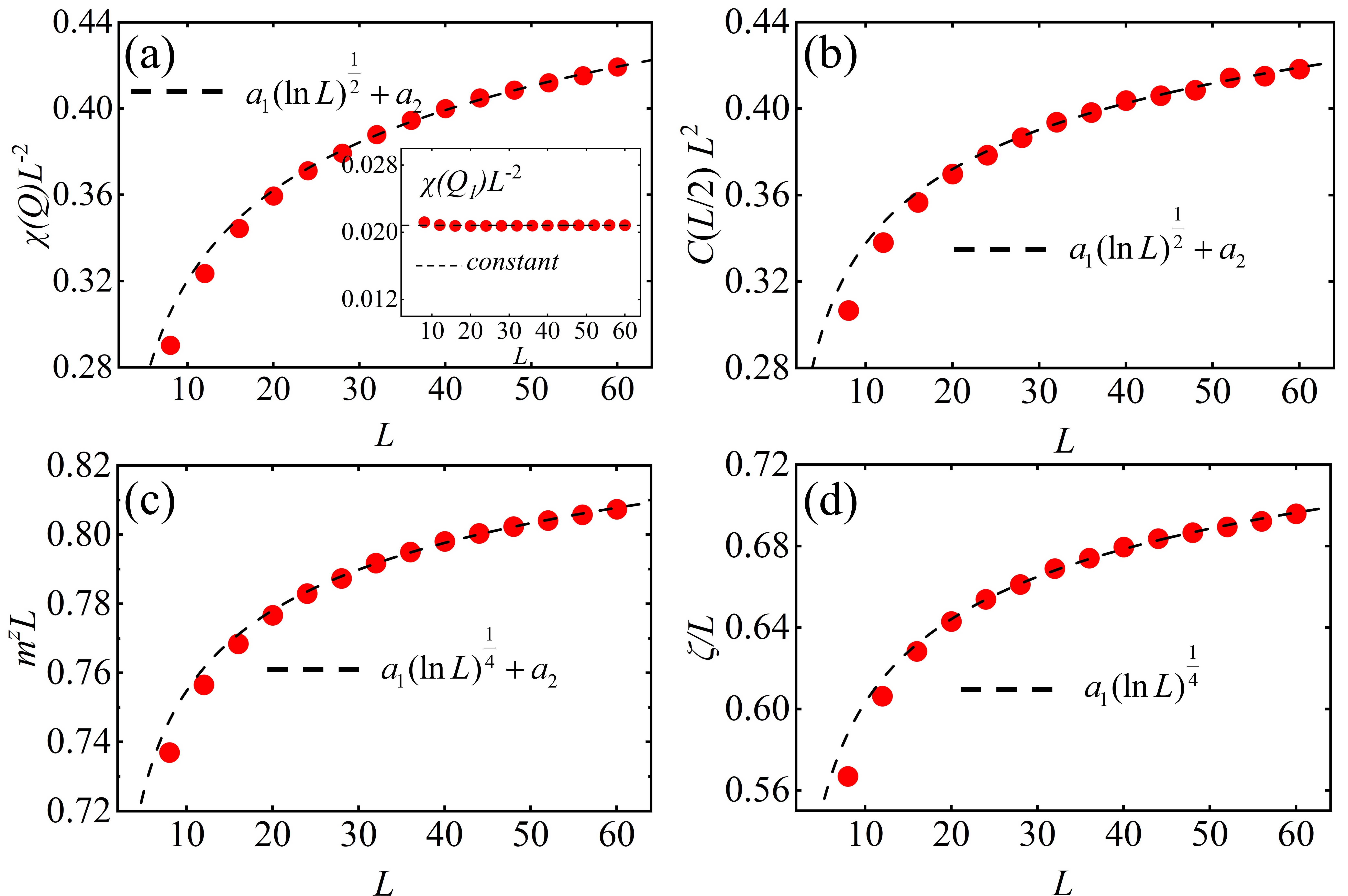}
    \caption{Bulk observables of the columnar-dimerized model at the bulk QCP. (a) Scaled staggered susceptibility $\chi(Q)L^{-2}$ versus $L$; the inset shows $\chi(Q_1)L^{-2}$, which approaches a finite constant without visible logarithmic corrections. (b) Scaled correlation function $C(L/2)L^{2}$ versus $L$. (c) Scaled magnetization $m^zL$ versus $L$. (d) Scaled correlation length $\xi/L$ versus $L$. The dotted lines denote least-squares fits; in (d), the fit is consistent with the predicted logarithmic finite-size exponent $\hat{\qq}=1/4$.}
    \label{fig:bulkQCP}
\end{figure}

\XY{Bulk critical behavior.} 
At the upper critical dimension of the $O(N)$ model, the correlation length, susceptibility, and magnetization exhibit power-law singularities with multiplicative logarithmic corrections~\cite{wegner1976critical,Amnon1983Nonlinear,Kenna2006scaling,Kenna2006Self-Consistent}
\begin{align}
\xi\sim \left | t \right |^{-\nu}\left | \ln \left | t \right | \right |^{\hat{\nu}},
\label{Eq:corlength-Fss-1}\\[1.0em]
\chi\sim \left | t \right |^{-\gamma}\left | \ln \left | t \right | \right |^{\hat{\gamma}},
\label{Eq:chi-Fss}\\[1.0em]
m\sim \left | t \right |^{\beta}\left | \ln \left | t \right | \right |^{\hat{\beta}},
\label{Eq:ms-Fss}
\end{align}
in which $t=g-g_c $, while at $t=0$, the correlation function obeys
\begin{equation}
\begin{array}{l}
\begin{aligned}
C(r)\sim r^{-(D-2+\eta)}(\ln r)^{\hat{\eta}}.
\end{aligned}
\end{array}
\label{Eq:cor-Fss}
\end{equation}
Here, the leading critical exponents take their mean-field values, $\nu=\frac{1}{2}$, $\gamma=1$, $\beta=\frac{1}{2}$ and $\eta=0$. The marginally irrelevant interaction generates multiplicative logarithmic corrections characterized by $\hat{\nu}=\frac{(N+2)}{2(N+8)}$, $\hat{\gamma}=\frac{(N+2)}{(N+8)}$, $\hat{\beta}=\frac{3}{(N+8)}$ and $\hat{\eta}=0$ for the $O(N)$ spin models~\cite{brezin1982investigation,brezin1976field,Kenna2006scaling,Kenna2006Self-Consistent}. In this framework, the finite-size scaling of the correlation length is governed by the logarithmic finite-size exponent
\begin{equation}
\begin{array}{l}
\begin{aligned}
\xi \sim L(\ln L)^{\hat{\qq}},
\end{aligned}
\end{array}
\label{Eq:corlength-Fss-2}
\end{equation}
with $\hat{\qq}=1/4$ for $O(N)$ models at the upper critical dimension~\cite{Kenna2006Self-Consistent,Kenna2013Anew,Kenna2014Fisher,brezin1982investigation}. Using Eqs.~\eqref{Eq:corlength-Fss-1}--\eqref{Eq:corlength-Fss-2}, the susceptibility, magnetization, and correlation function obey the following finite-size scaling forms at criticality
\begin{align}
\chi&\sim L^{\frac{\gamma}{\nu}}(\ln L)^{\hat{\gamma}-\frac{\gamma}{\nu}{(\hat{\nu}-\hat{\qq})}},
\label{Eq:susceptibility-Q-Fss}\\[1.0em]
m&\sim L^{-\frac{\beta}{\nu}}(\ln L)^{\hat{\beta}+\frac{\beta}{\nu}{(\hat{\nu}-\hat{\qq})}},
\label{Eq:magnetization-Q-Fss}\\[1.0em]
C(L/2)&\sim L^{-2-\eta}(\ln L)^{\hat{\eta}+(2-\eta)\hat{\qq}}.
\label{Eq:correlation-Q—Fss}
\end{align}
However, Lv et al. found that short-distance correlations and susceptibilities at non-ordering Fourier modes do not exhibit the leading multiplicative logarithmic enhancement in the classical 4D $O(N)$ models~\cite{Li2024Logarithmic,Lv2021Finite}. To account for the finite-size scaling of distance-dependent observables, they extended the standard free-energy density proposed by Kenna through the introduction of an additional term $\tilde{f_0}$~\cite{KENNA2004finite,Fang2021Logarithmic,Li2024Logarithmic,Lv2021Finite}
\begin{equation}
\begin{array}{l}
\begin{aligned}
f(t,h)=\;& L^{-4}\tilde{f_0}\!\left(tL^{y_t},\, hL^{y_h}\right)\\

\\

&+L^{-4}\tilde{f_1}\!\left(tL^{y_t}(\ln L)^{\hat{y}_t},\, hL^{y_h}(\ln L)^{\hat{y}_h}\right),
\end{aligned}
\end{array}
\label{Eq:Fss2}
\end{equation}
where the renormalization exponents $y_t=2$, $y_h=3$ and the logarithmic exponents $\hat{y}_t=(4-N)/(2N+16)$, $\hat{y}_h=1/4$. Thus, Eqs.~\eqref{Eq:susceptibility-Q-Fss}--\eqref{Eq:correlation-Q—Fss} take the forms
\begin{align}
\chi(Q)&= a_1 L^{2}(\ln L)^{\frac{1}{2}}+a_2L^{2},
\label{Eq:susceptibility-Q-Fss-1}\\[1.0em]
m&=a_1 L^{-1}(\ln L)^{\frac{1}{4}}+a_2 L^{-1},
\label{Eq:magnetization-Q-Fss-1}\\[1.0em]
C(L/2)&= a_1L^{-2}(\ln L)^{\frac{1}{2}}+a_2L^{-2}
\label{Eq:correlation-Q—Fss-1}
\end{align}
in which $a_1$ and $a_2$ are nonuniversal, observable-dependent constants.
 
We perform QMC simulations of the Hamiltonian in Eq.~\eqref{Eq:Hmlt} for system sizes up to $L=60$, and measure several bulk observables at the bulk QCP, including the staggered susceptibility $\chi(Q)$ ($Q=(\pi,\pi,\pi)$), susceptibility non-ordering Fourier modes $\chi(Q_1)$ ($Q_1=(\pi+2\pi/L,\pi,\pi)$), magnetization $m^z$, two-point correlation function $C(L/2)$, and correlation length $\xi$. As shown in the inset of Fig.~\ref{fig:bulkQCP}(a), $\chi(Q_1)L^{-2}$ approaches a finite nonzero constant and shows no visible logarithmic drift, consistent with the absence of a leading logarithmic enhancement for this non-ordering mode. In Figs.~\ref{fig:bulkQCP}(a)-\ref{fig:bulkQCP}(c), the scaled quantities $\chi(Q)L^{-2}$, $m^zL$, and $C(L/2)L^{2}$ show clear logarithmic increases with system size rather than approaching size-independent constants at the QCP. The scaling forms in Eqs.~\eqref{Eq:susceptibility-Q-Fss-1}--\eqref{Eq:correlation-Q—Fss-1}  are consistent with the numerical results for $\chi(Q)L^{-2}$, $m^zL$ and $C(L/2)L^{2}$. Details of the fitting procedure are provided in the SM~\cite{SM1}. Notably, if the $a_2$ correction term is ignored, the fit gives ${\hat{\eta}+(2-\eta)\hat{\qq}}=0.40(1) $ for a minimum fitting size of $L_{min}=24$, which is smaller than the theoretical value 0.5. A similar fitting behavior was observed in Refs.~\cite{Lv2021Finite,Fang2021Logarithmic,Lai1990Finite-size}, indicating that the additive nonlogarithmic contribution should be included to reconcile the finite-size data with the predicted asymptotic logarithmic exponent. However, higher-order two-loop logarithmic corrections $O(\frac{\ln (\ln L)}{\ln L})$ may arise in principle~\cite{KENNA2004finite}. Over the accessible range of system sizes, however, they are strongly entangled with the nonlogarithmic background contribution and cannot be resolved reliably. The numerical results for $\xi /L$ are shown in Fig.~\ref{fig:bulkQCP}(d). As expected, the correlation length obeys the scaling $\xi \sim L(\ln L)^{\frac{1}{4}}$ in the $(3+1)$D O(3) spin model.

\begin{figure}
    \centering
    \includegraphics[width=0.8\linewidth]{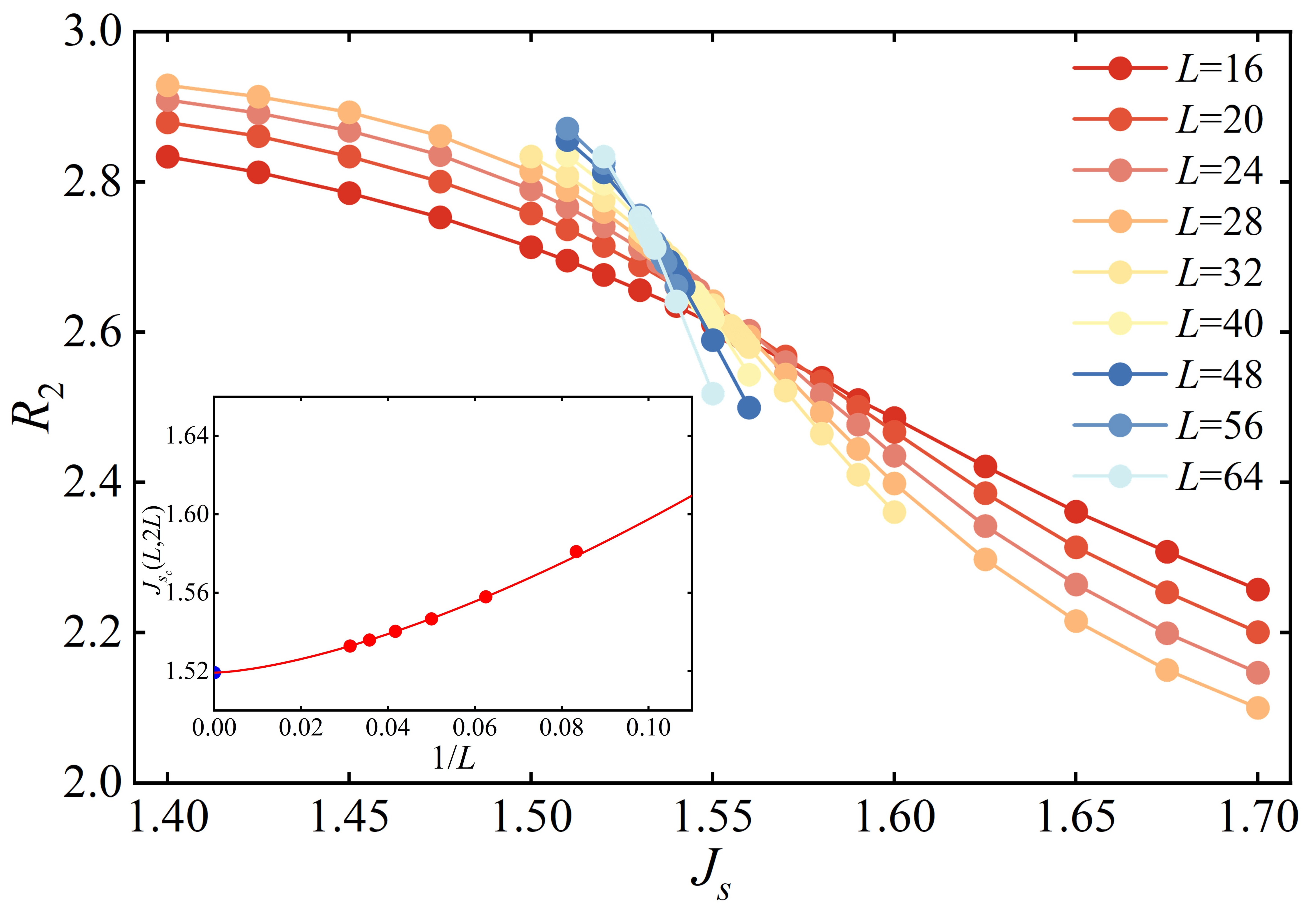}
    \caption{The surface Binder ratio $R_2$ versus $J_s$ for different system sizes at the bulk QCP. The inset shows the finite-size estimates $J_{s_c}(L)$ obtained from the crossings of curves for pairs of system sizes $L$ and $2L$, and the line represents the fitting function in Eq.~\eqref{Eq:extrapolation_Js} by neglecting the logarithmic correction. The blue symbol shows the extrapolated value in the thermodynamic limit.}
    \label{fig:speicalQCP}
\end{figure}

\XY{Special transition.} 
To locate the special transition, we tune the surface coupling $J_s$ while fixing $J_2/J_1$ at its bulk critical value. We impose open boundary conditions along the $[100]$ direction and periodic boundary conditions along the other two spatial directions. Fig.~\ref{fig:speicalQCP} shows $R_2$ as a function of $J_s$ for different system sizes. The curves exhibit a common crossing, which identifies the special transition separating the ordinary and extraordinary regimes. The critical coupling $J_{s_c}$ in the thermodynamic limit can be estimated using the standard $(L,2L)$ crossing analysis~\cite{Shao2016quantum,qin2015multiplicative}
\begin{equation}
\begin{array}{l}
J_{s_c}(L)=J_{s_c}+bL^{-y_{sp}-\omega}(\ln L)^{\hat{c}}.
\end{array}
\label{Eq:extrapolation_Js}
\end{equation}
Here, $y_{sp}$ is the scaling dimension of the relevant scaling field at the special transition. In $D=4-\epsilon$ dimensions, its $\epsilon$ expansion is~\cite{Parisen2021boundary,diehl1981field}
\begin{equation}
\begin{array}{l}
y_{sp}=1-\frac{N+2}{N+8}\epsilon+O(\epsilon^2).
\end{array}
\label{Eq:msz}
\end{equation}
The upper critical dimension corresponds to $\epsilon=0$, for which $y_{sp}=1$. Here, $\omega$ is the correction-to-scaling exponent, while $\hat{c}$ denotes the logarithmic correction exponent associated with the surface correction term, introduced in analogy with its bulk counterpart~\cite{qin2015multiplicative}. However, the available size window is not sufficient to stably resolve all logarithmic factors. We avoid fitting the data with the full multiplicative logarithmic form and instead adopt an effective algebraic scaling form obtained from Eq.~\eqref{Eq:extrapolation_Js} by dropping the logarithmic correction. In this effective fit, $\omega$ should be regarded as an effective exponent that absorbs crossover and other subleading finite-size effects over the accessible system-size range. This gives the thermodynamic-limit estimate $J_{s_c}=1.519(1)$, as shown in the inset of Fig.~\ref{fig:speicalQCP}.
\begin{figure}
    \centering
    \includegraphics[width=1.0\linewidth]{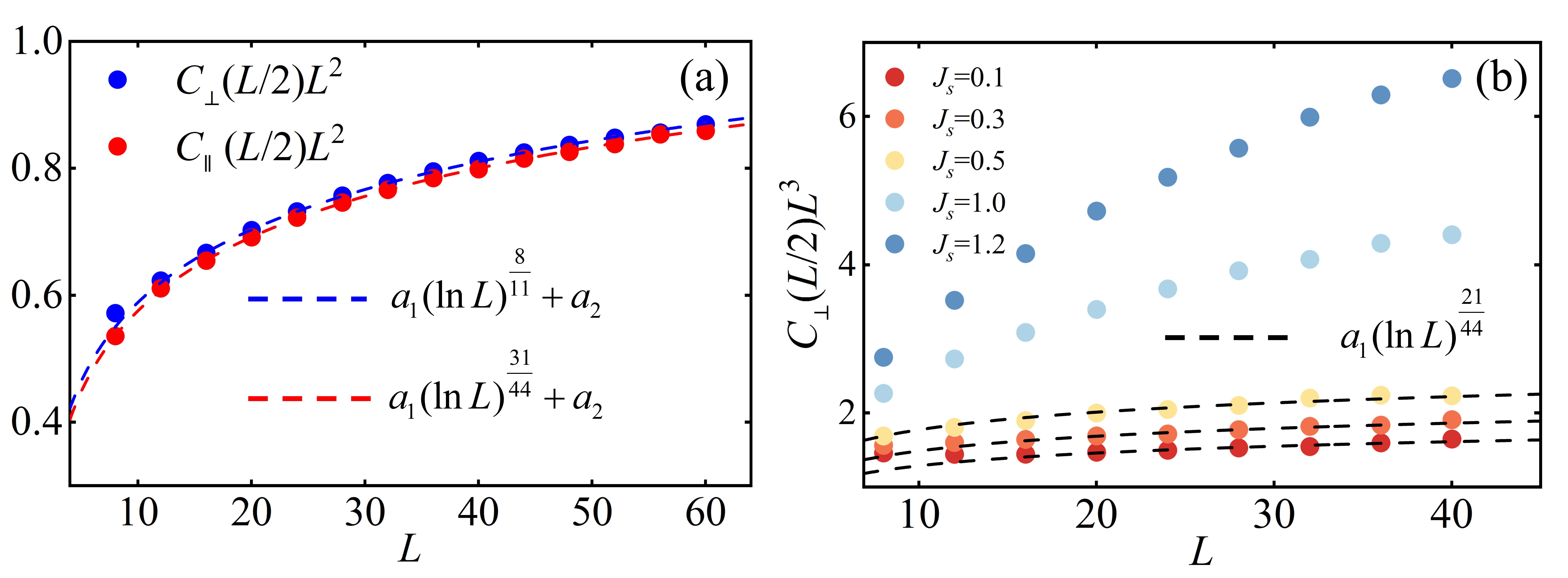}
    \caption{(a) The scaled correlation functions $C_{\parallel}(L/2)L^{2}$ and $C_{\perp}(L/2)L^{2}$ versus $L$ at the special transition. (b) The scaled correlation functions $C_{\perp}(L/2)L^{3}$ as a function of $L$ for several values of $J_s$ in the ordinary regime.}
    \label{fig:Q-cor-Fss}
\end{figure}

The two-point correlation functions are expected to exhibit the following scaling behavior at the critical point
\begin{align}
C_{\parallel}(r) &\sim r^{-(D-2+\eta_{\parallel})}
(\ln r)^{\hat{\eta}_{\parallel}},
\label{Eq:cor-parallel}\\[1.0em]
C_{\perp}(r) &\sim r^{-(D-2+\eta_{\perp})}
(\ln r)^{\hat{\eta}_{\perp}} .
\label{Eq:cor-perp}
\end{align}
Here, $\eta_{\parallel}$ and $\eta_{\perp}$ are the leading surface anomalous dimensions, while $\hat{\eta}_{\parallel}$ and $\hat{\eta}_{\perp}$ characterize the corresponding multiplicative logarithmic correction exponents. The RG analysis gives vanishing leading boundary anomalous dimensions at the special transition, $\eta_{\parallel}=\eta_{\perp}=0$. The nontrivial scaling information is contained in the logarithmic correction exponents, with $\hat{\eta}_{\parallel}=(N+2)/(N+8)$ and $\hat{\eta}_{\perp}=(N+2)/[2(N+8)]$ (see the SM~\cite{SM1} for the derivation). At the upper critical dimension, the finite-size correlation length follows the scaling form in Eq.~\eqref{Eq:corlength-Fss-2}, and the bulk logarithmic length correction can be incorporated into the boundary scaling forms. The finite-size logarithmic exponents acquire additional contributions from $\hat{\qq}$. We therefore obtain
\begin{align}
\hat{\eta}_{\parallel_Q}&=\hat{\eta}_{\parallel}+(1-\eta_\parallel)\hat{\qq},
\label{Eq:Q-cor-Fss-parallel}\\[1.0em]
\hat{\eta}_{\perp_Q}&=\hat{\eta}_{\perp}+(2-\eta_\perp)\hat{\qq}.
\label{Eq:Q_Fss}
\end{align}
Here, $\hat{\eta}_{\parallel_Q}$ and $\hat{\eta}_{\perp_Q}$ denote the finite-size logarithmic correction exponents. Thus, the finite-size scaling forms of Eqs.~\eqref{Eq:cor-parallel} and~\eqref{Eq:cor-perp} are
\begin{align}
C_{\parallel}(L/2) &\sim L^{-(D-2+\eta_\parallel)}(\ln L)^{\hat{\eta}_{\parallel_Q}}, 
\label{Eq:cor-Fss-parallel}\\[1.0em]
C_{\perp}(L/2)& \sim L^{-(D-2+\eta_\perp)}(\ln L)^{\hat{\eta}_{\perp_Q}}.
\label{Eq:cor-Fss-perp}
\end{align}
For the $D=4$ $O(3)$ model at the special transition, the above relations give
$\hat{\eta}_{\parallel_Q}=31/44$ and $\hat{\eta}_{\perp_Q}=8/11$. To test these nontrivial logarithmic corrections in finite-size scaling, we plot the scaled correlation functions 
$C_{\parallel}(L/2)L^{2}$ and $C_{\perp}(L/2)L^{2}$ at the special transition in Fig.~\ref{fig:Q-cor-Fss}(a). Both quantities exhibit a clear logarithmic increase with system size, instead of approaching size-independent constants. The two curves show very similar size dependence over the investigated range, consistent with the close theoretical values of $\hat{\eta}_{\parallel_Q}$ and $\hat{\eta}_{\perp_Q}$. We find that the scaling forms in Eqs.~\eqref{Eq:cor-Fss-parallel} and \eqref{Eq:cor-Fss-perp} with an additional nonlogarithmic background term $a_2L^{-2}$ describe the numerical data well. Details of the fitting procedure are provided in the SM~\cite{SM1}. These results provide direct numerical support for the predicted logarithmically corrected finite-size scaling at the special transition of the $(3+1)$D $\textrm{O}(3)$ QCP.

\XY{Ordinary transition.} 
For $J_s<J_{s_c}$, the system is in the ordinary boundary regime, where the boundary remains disordered and exhibits critical singularities only at the bulk QCP. The correlation functions follow the same logarithmically corrected forms as Eqs.~\eqref{Eq:cor-parallel} and~\eqref{Eq:cor-perp}, with exponents $\eta_{\parallel}=2$ and $\eta_{\perp}=1$ at the ordinary transition. The associated logarithmic correction exponents are $\hat{\eta}_{\parallel}=(N+2)/(N+8)$ and $\hat{\eta}_{\perp}=(N+2)/[2(N+8)]$~\cite{diehl1981field2} (see also the SM~\cite{SM1}). Consequently, the finite-size logarithmic corrections in Eqs.~\eqref{Eq:Q-cor-Fss-parallel} and~\eqref{Eq:Q_Fss} give $\hat{\eta}_{\parallel_Q}=9/44$ and $\hat{\eta}_{\perp_Q}=21/44$ for the $O(3)$ model. The rapid decay of $C_{\parallel}(L/2)$, however, makes it difficult to reliably extract its logarithmic correction exponent. We instead focus on $C_{\perp}(L/2)$. In Fig.~\ref{fig:Q-cor-Fss}(b), we plot $C_{\perp}(L/2)L^3$ for several values of $J_s$ in the ordinary regime. For sufficiently weak boundary couplings ($J_s\le 0.5$), which are well separated from the special transition, the data are compatible with the predicted logarithmic form $C_{\perp}(L/2)L^3\sim(\ln L)^{21/44}$. As $J_s$ approaches $J_{s_c}$, the proximity to the special transition produces stronger crossover corrections. These crossover effects obscure the asymptotic ordinary scaling behavior and make it difficult to resolve the predicted logarithmic correction exponent.

\begin{figure}
    \centering
    \includegraphics[width=1\linewidth]{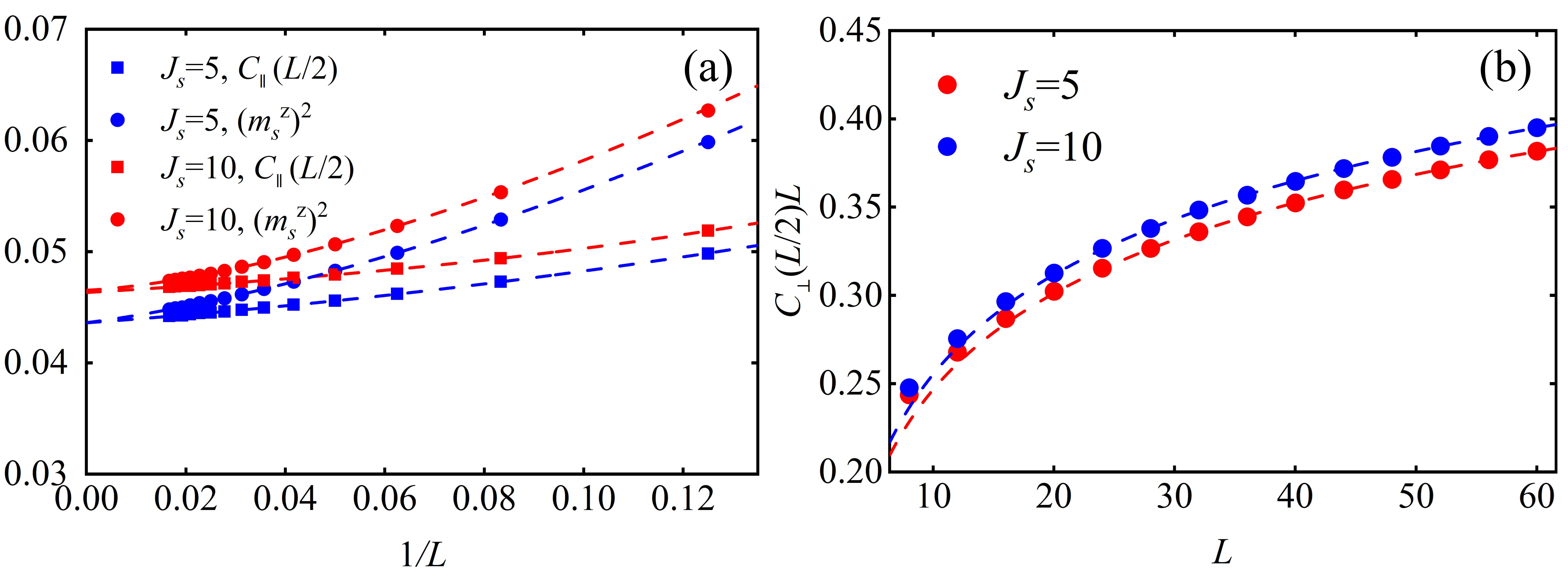}
    \caption{(a) Squared surface staggered magnetization ${m_s^z}^2$ and correlation function $C_{\parallel}(L/2)$ versus $1/L$ and (b) the scaled correlation function $C_{\perp}(L/2)L$ versus system size $L$ in the  extraordinary regime. The dotted lines represent the least-squares fits.}
    \label{fig:extraordinary}
\end{figure}

\XY{Extraordinary transition.} 
The regime $J_s>J_{s_c}$ corresponds to the extraordinary transition. Many of its universal properties are expected to coincide with those of the normal universality class~\cite{Bray1977Critical,Burkhardt1987Surface,Burkhardt1994Ordinary,Diehl1994Critical}. In this case, the boundary Goldstone modes effectively decouple from the bulk critical fluctuations and behave as free bosons (see, e.g., Ref.~\cite{Metlitski2022boundary}). The surface correlation function $C_{\parallel}(L/2)$ and the squared surface magnetization ${m_s^z}^2$ are expected to obey
\begin{align}
C_{\parallel}(L/2)=C_{\parallel}+a_1L^{-1}+a_2L^{-\omega}, 
\label{Eq:cor-extraordinary-parallel}\\[1.0em]
{m_s^z}^2(L)={m_s^z}^2+a_1L^{-1}+a_2L^{-\omega} , 
\label{Eq:cor-extraordinary-ms}
\end{align}
where $\omega$ denotes an effective subleading correction exponent. Fitting the numerical data using Eqs.~\eqref{Eq:cor-extraordinary-parallel} and~\eqref{Eq:cor-extraordinary-ms}, we obtain the results shown in  Fig.~\ref{fig:extraordinary}(a). For both $J_s=5$ and $J_s=10$, which are well inside the extraordinary regime, the extrapolated values of $C_{\parallel}$ and ${m_s^z}^2$ agree within error bars, providing evidence for long-range surface magnetic order. A simple estimate of the leading finite-size scaling gives $C_{\perp}(L/2)\sim m_sm(L)$, where $m_s\neq0$ is the surface order parameter and $m$ is the induced bulk order-parameter profile. Using the scaling form of $m(L)$, we obtain $C_{\perp}(L/2)\sim L^{-1}(\ln L)^{1/4}$, i.e., $\hat{\eta}_{\perp_Q}=1/4$. Fig.~\ref{fig:extraordinary}(b) shows the rescaled quantity $C_{\perp}(L/2)L$ as a function of $L$ for $J_s=5$ and $J_s=10$. While the numerical results are consistent with the leading power law $L^{-1}$, $C_{\perp}(L/2)L$ exhibits a clear logarithmic enhancement. Fitting the data yields an effective logarithmic exponent $\hat{\eta}_{\perp Q}=0.757(4)$, which is substantially larger than the simple theoretical estimate. Related challenges in interpreting numerical scaling behaviors have also been mentioned in studies of extraordinary surface criticality in (2+1)D quantum magnets~\cite{Ding2018engineering,Wang2022bulk,Wang2024surface,Wang2023extraordinary}. In the present $(3+1)$D system, additional subtleties may arise from the fact that the bulk critical point lies at its upper critical dimension. A more complete theoretical understanding of this regime is left for future work.

\XY{Summary and outlook.} In summary, we have studied $\textrm{O}(N)$ boundary criticality at the bulk upper critical dimension. Using large-scale QMC simulations and boundary RG analysis, we showed that the logarithmically slow running of marginally irrelevant bulk interactions generates multiplicative logarithmic corrections to boundary correlations. At the ordinary and special transitions, we determined the corresponding logarithmic correction exponents and incorporated the logarithmic exponent $\hat{\qq}$ into the finite-size scaling forms. The resulting $\hat{\qq}$-dependent scaling forms are quantitatively supported by our QMC results. In the extraordinary regime, we found long-range surface magnetic order and a logarithmically enhanced surface-bulk correlation.

Our work suggests several directions for future study. Since the RG analysis applies to general $\textrm{O}(N)$ models, the predicted $N$ dependence of the logarithmic boundary exponents could be tested in other $\textrm{O}(N)$ systems and different microscopic realizations. Beyond conventional correlation functions, it would be interesting to explore how multiplicative logarithmic corrections manifest themselves in nonlocal observables associated with boundary criticality, such as disorder operators~\cite{liu2024measuring}. Another direction is to investigate boundary criticality at unconventional quantum critical points involving marginally irrelevant bulk interactions, such as the continuous Mott transition~\cite{podolsky2009mott}. Finally, recent realizations of the three-dimensional fermionic Hubbard model in ultracold-atom experiments~\cite{shao2024antiferromagnetic,Wang2025Homogeneous} provide a promising setting for studying boundary critical phenomena in quantum simulators. Whether the logarithmic boundary scaling identified here can be observed in such systems remains an interesting question for future work.

\XY{Acknowledgments.}
X.L. and X.W. contributed equally in this work. X.L. and D.X.Y. were supported by NKRDPC-2022YFA1402802, NSFC-92165204, NSFC-12494591, Guangdong Provincial Key Laboratory of Magnetoelectric Physics and Devices (2022B1212010008), Guangdong Fundamental Research Center for Magnetoelectric Physics (2024B0303390001), and Guangdong Provincial Quantum Science Strategic Initiative (GDZX2401010). X.W. was supported by the Simons Investigator Grant (566116) awarded to S. Ryu. Z.L. is supported by the China Postdoctoral Science Foundation under Grant No.2024M762935. Z.W. is supported by the China Postdoctoral Science Foundation under Grant No.~2024M752898. Z. Y. is supported by the the Scientific Research Project (Grant No. WU2025B011), Feng-Ying Career Development Chair Fund and start-up funding of Westlake University.
\bibliography{Ref}
\clearpage
\newpage
\onecolumngrid
\setcounter{page}{1}

\appendix
\begin{center}
\textbf{\large Supplemental Material for ``Logarithmic corrections to bulk and surface criticality in a three-dimensional quantum Heisenberg antiferromagnet''}\\
\end{center}

\begin{center}
{
Xuyang Liang,$^{1}$
Xiao-Chuan Wu,$^{2}$
Zenan Liu,$^{3,4}$
Zhe Wang,$^{3,4}$
Zheng Yan,$^{3,4}$
and Dao-Xin Yao$^{1}$
}\par
\vspace{0.4em}

{\itshape
$^{1}$Guangdong Provincial Key Laboratory of Magnetoelectric Physics and Devices,\\
State Key Laboratory of Optoelectronic Materials and Technologies,\\
Institute of Neutron Science and Technology, School of Physics,\\
Sun Yat-Sen University, Guangzhou, 510275, China\\[0.25em]
$^{2}$Department of Physics, Princeton University, Princeton, NJ 08544, USA\\
$^{3}$Department of Physics, School of Science and Research Center for\\
Industries of the Future, Westlake University, Hangzhou 310030, China\\
$^{4}$Institute of Natural Sciences, Westlake Institute for Advanced Study, Hangzhou 310024, China
}\par
\end{center}

\section{Critical exponents for logarithmic corrections}

In this section, we determine several critical exponents associated with logarithmic corrections in the boundary critical behavior of the $\textrm{O}(N)$ vector model at its upper critical dimension. To the best of our knowledge, explicit results for $\hat{\eta}_{\Vert}$ and $\hat{\eta}_{\perp}$ at the special transition, as given in Eq.~\eqref{eq:_special_anom_dim}, as well as the finite-size scaling formulas that include the effect of the enigmatic exponent $\hat{\qq}$, such as Eqs.~\eqref{eq:_ordinary_suscept_FSS} and~\eqref{eq:_special_suscept_FSS}, are not available in the existing literature.

Our analysis is based on the $\textrm{O}(N)$ vector model defined on the semi-infinite manifold $\{r^{\mu}=(\boldsymbol{r}^{\Vert},r^{\perp})\in\mathbb{R}^{4}|r^{\perp}\geq0\}$
\begin{flalign}
\mathcal{S}[\boldsymbol{\phi}]=\int\textrm{d}^{3}\boldsymbol{r}^{\Vert}\left(\tilde{c}|\boldsymbol{\phi}|^{2}+\tilde{h}\cdot\boldsymbol{\phi}+\int^{+\infty}_{0}\textrm{d}r^{\perp}\left(\frac{1}{2}|\partial\boldsymbol{\phi}|^{2}+t|\boldsymbol{\phi}|^{2}+u|\boldsymbol{\phi}|^{4}+h\cdot\boldsymbol{\phi}\right)+\ldots\right),
\end{flalign}
where $\boldsymbol{\phi}$ denotes the $\textrm{O}(N)$ order parameter. The bulk critical exponents take their mean-field values,
\begin{flalign}
\nu=\frac{1}{2},\qquad\eta=0,\qquad\alpha=0,\qquad\beta=\frac{1}{2},\qquad\gamma=1,\qquad\delta=3,
\end{flalign}
which can be found in textbooks such as Ref.~\cite{Cardy1996Book}. In addition, the marginally irrelevant coupling generates multiplicative logarithmic corrections characterized by the exponents~\cite{Kenna2006scaling,Kenna2006Self-Consistent}
\begin{flalign}
\hat{\nu}=\frac{N+2}{2(N+8)},\quad\hat{\eta}=0,\quad\hat{\alpha}=\frac{4-N}{N+8},\quad\hat{\beta}=\frac{3}{N+8},\quad\hat{\gamma}=\frac{N+2}{N+8},\quad\hat{\delta}=\frac{1}{3}.
\end{flalign}
These exponents can be derived from the RG running of the bulk couplings $u(\ell)\sim\ell^{\hat{\lambda}_{u}}$, $h(\ell)\sim e^{\lambda_{h}\ell}\ell^{\hat{\lambda}_{h}}$, and $t(\ell)\sim e^{\lambda_{t}\ell}\ell^{\hat{\lambda}_{t}}$, where $\ell$ denotes the scale parameter (see, e.g.~\cite{ruiz2017revisiting}). The corresponding RG exponents are
\begin{flalign}
\hat{\lambda}_{u}=-1,\qquad\lambda_{h}=3,\qquad\hat{\lambda}_{h}=0,\qquad\lambda_{t}=2,\qquad\hat{\lambda}_{t}=-\frac{N+2}{N+8}.
\end{flalign}
In the boundary critical problem, the two-point function of the order parameter exhibits the asymptotic behavior
\begin{flalign}
C(\boldsymbol{r}^{\Vert}_{1}-\boldsymbol{r}^{\Vert}_{2};r^{\perp}_{1},r^{\perp}_{2})\sim\begin{cases}
|\boldsymbol{r}^{\Vert}_{1}-\boldsymbol{r}^{\Vert}_{2}|^{-(D-2+\eta_{\parallel})}(\log|\boldsymbol{r}^{\Vert}_{1}-\boldsymbol{r}^{\Vert}_{2}|)^{\hat{\eta}_{\Vert}} & |\boldsymbol{r}^{\Vert}_{1}-\boldsymbol{r}^{\Vert}_{2}|\rightarrow\infty\textrm{ and }0<r^{\perp}_{1},r^{\perp}_{2}<\infty\\
|r^{\perp}_{1}|^{-(D-2+\eta_{\perp})}(\log|r^{\perp}_{1}|)^{\hat{\eta}_{\perp}} & r^{\perp}_{1}\rightarrow\infty\textrm{ and }0<|\boldsymbol{r}^{\Vert}_{1}-\boldsymbol{r}^{\Vert}_{2}|,r^{\perp}_{2}<\infty\\
|r_{1}-r_{2}|^{-(D-2+\eta)}(\log|r_{1}-r_{2}|)^{\hat{\eta}} & |\boldsymbol{r}^{\Vert}_{1}-\boldsymbol{r}^{\Vert}_{2}|,r^{\perp}_{1},r^{\perp}_{2}\rightarrow\infty
\end{cases},
\end{flalign}
where $D=4$. To determine the values of $\eta_{\parallel},\eta_{\perp},\hat{\eta}_{\Vert}$ and $\hat{\eta}_{\perp}$, one must analyze the RG flow of the boundary couplings.

\subsection{Ordinary transition}

The ordinary transition corresponds to $\tilde{c}\rightarrow+\infty$. Following the operator product expansion (OPE) approach~\cite{Cardy1996Book}, the boundary RG flow of the (rescaled dimensionless) coupling constant is 
\begin{flalign}
\frac{\textrm{d}\tilde{h}}{\textrm{d}\ell}=\tilde{h}+4(N+2)\tilde{h}u(\ell),
\end{flalign}
which can be solved as
\begin{flalign}
\tilde{h}(\ell)=\tilde{h}(0)e^{\ell}(1+8(N+8)u(0)\ell)^{\frac{N+2}{2(N+8)}}.
\end{flalign}
We therefore identify the large-$\ell$  behavior $\tilde{h}(\ell)\sim e^{\lambda_{\tilde{h}}\ell}\ell^{\hat{\lambda}_{\tilde{h}}}$ with 
\begin{flalign}
\lambda_{\tilde{h}}=1,\qquad\hat{\lambda}_{\tilde{h}}=\frac{N+2}{2(N+8)}.
\end{flalign}
The boundary anomalous dimensions then follow directly from these RG exponents
\begin{flalign}
\eta_{\parallel}&=D-2\lambda_{\tilde{h}}=2,\qquad\hat{\eta}_{\Vert}=2\hat{\lambda}_{\tilde{h}}=\frac{N+2}{N+8},\nonumber\\\eta_{\perp}&=\frac{D+\eta}{2}-\lambda_{\tilde{h}}=1,\qquad\hat{\eta}_{\perp}=\hat{\lambda}_{h}+\hat{\lambda}_{\tilde{h}}=\frac{N+2}{2(N+8)}.
\label{eq:_ordinary_anom_dim}
\end{flalign}

Upon tuning the bulk coupling $t$, while keeping $h=\tilde{h}=0$, the surface susceptibilities (at zero momentum) exhibit the scaling behaviors
\begin{flalign}
\chi_{1}\sim\frac{1}{|t|^{\gamma_{1}}}\left(\log\frac{1}{|t|}\right)^{\hat{\gamma}_{1}},\qquad\chi_{11}\sim\frac{1}{|t|^{\gamma_{11}}}\left(\log\frac{1}{|t|}\right)^{\hat{\gamma}_{11}}.
\label{eq:_surface_suscept}
\end{flalign}
The corresponding critical exponents are given by
\begin{flalign}
\gamma_{11}&=\nu(1-\eta_{\parallel})=-\frac{1}{2},\qquad\hat{\gamma}_{11}=\hat{\nu}(1-\eta_{\parallel})+\hat{\eta}_{\Vert}=\frac{N+2}{2(N+8)},\nonumber\\\gamma_{1}&=\nu(2-\eta_{\perp})=\frac{1}{2},\qquad\hat{\gamma}_{1}=\hat{\nu}(2-\eta_{\perp})+\hat{\eta}_{\perp}=\frac{N+2}{N+8}.
\label{eq:_ordinary_suscept}
\end{flalign}
Equations~\eqref{eq:_ordinary_anom_dim} and~\eqref{eq:_ordinary_suscept} reproduce the results previously derived in Ref.~\cite{diehl1981field2}. Here, we further analyze finite-size scaling. Particular attention should be paid to the fact that the correlation length scales as $\xi_{L}\sim L(\log L)^{\hat{\text{ϙ}}}$, where $\hat{\text{ϙ}}=1/4$ is the ``enigmatic exponent'' (see Ref.~\cite{kenna2015scaling} and references therein). Substituting $t\sim L^{-1/\nu}(\log L)^{(\hat{\nu}-\hat{\text{ϙ}})/\nu}$, we obtain the finite-size behavior of the surface susceptibilities
\begin{flalign}
\chi_{1,L}&\sim L^{\gamma_{1}/\nu}(\log L)^{\hat{\gamma}_{1}-\gamma_{1}(\hat{\nu}-\hat{\text{ϙ}})/\nu}=L(\log L)^{\frac{3(N+4)}{4(N+8)}},\nonumber\\\chi_{11,L}&\sim L^{\gamma_{11}/\nu}(\log L)^{\hat{\gamma}_{11}-\gamma_{11}(\hat{\nu}-\hat{\text{ϙ}})/\nu}=L^{-1}(\log L)^{\frac{3N}{4(N+8)}}.
\label{eq:_ordinary_suscept_FSS}
\end{flalign}

\subsection{Special transition}
The special transition corresponds to a boundary critical point at $\tilde{c}=0$. We again use the real-space OPE approach~\cite{Cardy1996Book}, which gives the RG flow equations for the boundary coupling constants 
\begin{flalign}
\frac{\textrm{d}\tilde{h}}{\textrm{d}\ell}&=2\tilde{h}+4(N+2)\tilde{h}u(\ell)\nonumber\\\frac{\textrm{d}\tilde{c}}{\textrm{d}\ell}&=\tilde{c}-8(N+2)\tilde{c}u(\ell).
\end{flalign}
In the presence of the logarithmically slow running of $u(\ell)$, these equations are solved by
\begin{flalign}
\tilde{h}(\ell)&=\tilde{h}(0)e^{2\ell}(1+8(N+8)u(0)\ell)^{\frac{N+2}{2(N+8)}},\nonumber\\\tilde{c}(\ell)&=\tilde{c}(0)e^{\ell}(1+8(N+8)u(0)\ell)^{-\frac{N+2}{N+8}}.
\end{flalign}
Thus, the large-$\ell$ behavior can be written as $\tilde{h}(\ell)\sim e^{\lambda_{\tilde{h}}\ell}\ell^{\hat{\lambda}_{\tilde{h}}}$ and $\tilde{c}(\ell)\sim e^{\lambda_{\tilde{c}}\ell}\ell^{\hat{\lambda}_{\tilde{c}}}$ with 
\begin{flalign}
\lambda_{\tilde{h}}&=2,\qquad\hat{\lambda}_{\tilde{h}}=\frac{N+2}{2(N+8)},\nonumber\\\lambda_{\tilde{c}}&=1,\qquad\hat{\lambda}_{\tilde{c}}=-\frac{N+2}{N+8}.
\end{flalign}
These RG eigenvalues determine the boundary anomalous dimensions
\begin{flalign}
\eta_{\parallel}&=D-2\lambda_{\tilde{h}}=0,\qquad\hat{\eta}_{\Vert}=2\hat{\lambda}_{\tilde{h}}=\frac{N+2}{N+8},\nonumber\\\eta_{\perp}&=\frac{D+\eta}{2}-\lambda_{\tilde{h}}=0,\qquad\hat{\eta}_{\perp}=\hat{\lambda}_{h}+\hat{\lambda}_{\tilde{h}}=\frac{N+2}{2(N+8)}.
\label{eq:_special_anom_dim}
\end{flalign}
The critical exponents governing the surface susceptibilities in Eq.~\eqref{eq:_surface_suscept} are therefore
\begin{flalign}
\gamma_{11}&=\nu(1-\eta_{\parallel})=\frac{1}{2},\qquad\hat{\gamma}_{11}=\hat{\nu}(1-\eta_{\parallel})+\hat{\eta}_{\Vert}=\frac{3(N+2)}{2(N+8)},\nonumber\\\gamma_{1}&=\nu(2-\eta_{\perp})=1,\qquad\hat{\gamma}_{1}=\hat{\nu}(2-\eta_{\perp})+\hat{\eta}_{\perp}=\frac{3(N+2)}{2(N+8)}.
\label{eq:_special_suscept}
\end{flalign}
These results are consistent with Ref.~\cite{diehl1983multicritical}. The finite-size scaling analysis proceeds in the same way as for the ordinary transition. Including the same finite-size correction associated with the enigmatic exponent $\hat{\text{ϙ}}$, we obtain 
\begin{flalign}
\chi_{1,L}&\sim L^{\gamma_{1}/\nu}(\log L)^{\hat{\gamma}_{1}-\gamma_{1}(\hat{\nu}-\hat{\text{ϙ}})/\nu}=L^{2}(\log L)^{\frac{N+5}{N+8}},\nonumber\\\chi_{11,L}&\sim L^{\gamma_{11}/\nu}(\log L)^{\hat{\gamma}_{11}-\gamma_{11}(\hat{\nu}-\hat{\text{ϙ}})/\nu}=L(\log L)^{\frac{5N+16}{4(N+8)}}.
\label{eq:_special_suscept_FSS}
\end{flalign}


\section{Finite-size scaling of the bulk observables at the bulk QCP} 
In this section, we show the details of the fitting results of staggered magnetic susceptibility $\chi(Q)$, magnetization $m^z$, and correlation function $C(L/2)$. These quantities obey the following finite-size scaling forms. 

\begin{align}
\chi(Q)&= a_1L^{\frac{\gamma}{\nu}}(\ln L)^{\hat{\gamma}-\frac{\gamma}{\nu}{(\hat{\nu}-\hat{\qq})}}+a_2L^{\frac{\gamma}{\nu}},
\label{Eq:susceptibility-Q-Fss-2}\\[1.0em]
m^z&=a_1 L^{-\frac{\beta}{\nu}}(\ln L)^{\hat{\beta}+\frac{\beta}{\nu}{(\hat{\nu}-\hat{\qq})}}+a_2 L^{-\frac{\beta}{\nu}},
\label{Eq:magnetization-Q-Fss-2}\\[1.0em]
C(L/2) &= a_1L^{-2-\eta}(\ln L)^{\hat{\eta}+(2-\eta)\hat{\qq}}+a_2L^{-2-\eta}.
\label{Eq:correlation-Q—Fss-2}
\end{align}
The fit results are summarized in Tables~\ref{tab:chi}--\ref{tab:correlation} for the three-dimensional columnar-dimerized quantum Heisenberg model. First, we neglect the $a_2$ term and fit the data using only the leading contribution. We either fix the leading power-law exponents ${\frac{\gamma}{\nu}}$, ${\frac{\beta}{\nu}}$, ${-2-\eta}$ to their theoretical values and treat the corresponding logarithmic correction exponents as free fitting parameters, or fix the logarithmic correction exponents and determine the power-law exponents from the fits. The effective exponents exhibit noticeable deviations from the theoretical values, and these deviations do not systematically decrease as smaller system sizes are excluded, indicating appreciable subleading finite-size corrections. We therefore include the $a_2$ correction terms, fix all critical exponents to their theoretical values, and treat $a_1$ and $a_2$ as free fitting parameters. The fit quality improves as $L_{\min}$ is increased. These results indicate that the numerical data are consistent with the predicted scaling forms once the subleading finite-size corrections are included.

\begin{table*}[b]
\setlength{\tabcolsep}{8pt}
\caption{Fit to the finite-size staggered magnetic susceptibility $\chi(Q)$ data obtained from quantum Monte Carlo simulations at the bulk QCP based on Eq.~\eqref{Eq:susceptibility-Q-Fss-2}.}
\begin{tabular}{ccccccc}

\hline\hline
      $L_{min}$                       & ${\gamma}/{\nu}$    & ${\hat{\gamma}-\frac{\gamma}{\nu}{(\hat{\nu}-\hat{\qq})}}$     & $a_1$    &$a_2$      &$\chi^2/d.o.f$ \\ \hline
\\
                   28                 & 2           & 0.470(5)         &0.216(2)    &0                &47.3/7    \\      
\\
                   32                 & 2           & 0.449(6)         &0.222(2)   &0                &14.3/6    \\
\\
                   36                 & 2           & 0.437(8)        &0.226(3)     &0                &7.0/5   \\
\\
                   40                 & 2           & 0.425(10)         &0.230(4)    &0                &3.5/4  \\
\\
                   28                & 1.993(2)           & 0.5             &0.213(1)    &0                &36.0/7   \\  
\\
                   32                 & 1.989(2)           & 0.5             &0.216(2)    &0                &16.4/6    \\
\\
                   36                 & 1.988(2)           & 0.5             &0.217(2)    &0                &15.5/5   \\
\\  
                   40                 & 1.981(3)           & 0.5             &0.224(3)    &0                &3.4/4  \\
\\
                   28                 & 2           & 0.5              &0.196(2)    &0.023(4)         &48.5/7  \\
\\
                   32                 & 2           & 0.5              &0.187(3)    &0.041(5)         &15.0/6   \\
\\
                   36                 & 2           & 0.5              &0.182(3)    &0.051(6)        &7.2/5  \\
\\
                   40                 & 2           & 0.5              &0.176(5)    &0.061(9)        &3.6/4  \\
 \\ \hline\hline
\label{tab:chi}
\end{tabular}
\end{table*}

\begin{table*}[b]
\setlength{\tabcolsep}{8pt}
\caption{Fit to the finite-size magnetization $m^z$ data obtained from quantum Monte Carlo simulations at the bulk QCP based on Eq.~\eqref{Eq:magnetization-Q-Fss-2}.}
\begin{tabular}{ccccccc}

\hline\hline
      $L_{min}$                       & ${\beta}/{\nu}$    & ${\hat{\beta}+\frac{\beta}{\nu}{(\hat{\nu}-\hat{\qq})}}$     & $a_1$    &$a_2$      &$\chi^2/d.o.f$ \\ \hline
\\
                   28                 & 1                  & 0.121(1)         &0.681(1)    &0                  &42.4/7    \\      
\\
                   32                 & 1                  & 0.117(2)         &0.685(2)     &0                 &14.9/6    \\
\\
                   36                 & 1                   & 0.115(2)         &0.686(2)     &0                 &11.6/5   \\
\\ 
                   40                 & 1                  & 0.110(3)         &0.692(3)     &0                 &0.5/4  \\
\\
                   28                 & 1.0350(3)          & 0.25             &0.655(1)    &0                &18.4/7   \\      
\\
                   32                 & 1.0353(4)           & 0.25             &0.656(1)    &0                 &16.0/6    \\
\\
                   36                 & 1.0353(4)           & 0.25              &0.656(1)    &0                &16.0/5   \\
\\  
                   40                 & 1.0364(6)           & 0.25              &0.659(2)    &0                &9.4/4  \\
\\
                   28                 & 1                 &0.25                     &0.278(3)    &0.413(4)         &48.3/7  \\
\\
                   32                 & 1                 & 0.25                     &0.268(3)    &0.426(4)         &16.8/6  \\
\\ 
                   36                 & 1                 & 0.25                      &0.264(4)    &0.431(5)        &12.8/5 \\
\\ 
                   40                 & 1                 &0.25                      &0.251(6)    &0.450(8)         &0.6/4 \\
 \\ \hline\hline
\label{tab:magnetization}
\end{tabular}
\end{table*}

\begin{table*}[t]
\setlength{\tabcolsep}{8pt}
\caption{Fit to the finite-size staggered correlation function $C(L/2)$ data obtained from quantum Monte Carlo simulations at the bulk QCP based on Eq.~\eqref{Eq:correlation-Q—Fss-2}.}
\begin{tabular}{ccccccc}

\hline\hline
      $L_{min}$                       & $-2-\eta$    & ${\hat{\eta}+(2-\eta)\hat{\qq}}$     & $a_1$    &$a_2$      &$\chi^2/d.o.f$ \\ \hline
\\
                   16                 & -2                 & 0.421(5)          & 0.232(2)   &0                &21.3/10    \\      
\\
                   20                 & -2                 & 0.406(7)          &0.237(2)     &0                &11.9/9    \\
\\
                   24                 & -2                 & 0.40(1)          &0.238(3)     &0                &11.4/8    \\
\\
                   16                 & -2.022(2)           & 0.5             &0.228(1)    &0                &27.0/10   \\      
\\
                   20                 & -2.028(3)           & 0.5             &0.233(2)    &0                &13.3/9    \\
\\
                   24                 & -2.029(3)           & 0.5             &0.233(3)    &0                &13.1/8   \\
\\
                   16                 & -2           & 0.5                    &0.179(9)    &0.060(4)         &24.3/10  \\
\\
                   20                 & -2           & 0.5                    &0.171(3)    &0.074(6)         &12.9/9   \\
\\
                   24                 & -2           & 0.5                    &0.169(4)    &0.078(8)        &12.2/8  \\

 \\ \hline\hline
\label{tab:correlation}
\end{tabular}
\end{table*}

\section{Finite-size scaling of the parallel and perpendicular correlation functions at the special transition} 
In this section, we show the details of the fitting results of $C_{\parallel}(L/2)$ and $C_{\perp}(L/2)$. The quantities $C_{\parallel}(L/2)$ and $C_{\perp}(L/2)$ obey the following finite-size scaling forms
\begin{align}
C_{\parallel}(L/2)&= a_1L^{-2-\eta_\parallel}(\ln L)^{\hat{\eta}_{\parallel_Q}}+a_2L^{-2-\eta_\parallel},
\label{Eq:cor-Fss-parallel-1}\\[1.0em]
C_{\perp}(L/2)&= a_1L^{-2-\eta_{\perp}}(\ln L)^{\hat{\eta}_{\perp_Q}}+a_2L^{-2-\eta_{\perp}}.
\label{Eq:cor-Fss-perp-1}
\end{align}
The fit results are summarized in Tables~\ref{tab:C_parallel(L/2)} and~\ref{tab:C_perp(L/2)} for the three-dimensional columnar-dimerized quantum Heisenberg model. First, we omit the $a_2$ term and fit the data with the leading contribution. In this analysis, we either fix the leading boundary anomalous dimensions $\eta_\parallel$ and $\eta_\perp$, or fix the logarithmic correction exponents ${\hat{\eta}_{\parallel_Q}}$ and ${\hat{\eta}_{\perp_Q}}$, and then determine the remaining exponent from the fit. The fitted values of ${\hat{\eta}_{\parallel_Q}}$ and ${\hat{\eta}_{\perp_Q}}$ are found to be close to the theoretical prediction 31/44 and 8/11 with $L_{min}=28$, respectively. 
Conversely, when ${\hat{\eta}_{\parallel_Q}}$ and ${\hat{\eta}_{\perp_Q}}$  are fixed to their theoretical values, the fitted power-law exponents $-2-\eta_\parallel$ and $-2-\eta_\perp$ are also consistent with the expected values. We then include the $a_2$ correction term to improve the stability of the fits. In this case, all critical exponents are fixed to their theoretical values, and $a_1$ and $a_2$ are left as free fitting parameters. This procedure gives stable fits for all fitting windows with $L_{min}\geq20$, indicating that the numerical data are consistent with the predicted scaling forms.

\begin{table*}[b]
\setlength{\tabcolsep}{8pt}
\caption{Fit to the finite-size parallel correlation function $C_\parallel(L/2)$ data obtained from quantum Monte Carlo simulations at the special transition based on Eq.~\eqref{Eq:cor-Fss-parallel-1}.}
\begin{tabular}{ccccccc}

\hline\hline
      $L_{min}$                       & $-2-\eta_\parallel$    & $\hat{\eta}_{\parallel_Q}$     & $a_1$    &$a_2$      &$\chi^2/d.o.f$ \\ \hline
\\
                   12                 & -2                 & 0.688(4)           & 0.326(2)   &0                &12.9/11    \\    
\\
                   20                 & -2                  & 0.694(8)           &0.324(4)     &0                &5.8/9    \\
\\
                   28                 & -2                  & 0.69(2)           &0.324(6)     &0                 &4.7/7   \\
\\
                   12                 & -2.005(2)           & 31/44             &0.324(2)    &0                &12.4/11   \\    
\\
                   20                 & -2.001(3)           & 31/44             &0.321(3)    &0                &5.6/9   \\
\\
                   28                 & -2.003(4)           & 31/44             &0.323(4)    &0                &4.8/7   \\
\\
                   12                 & -2                  & 31/44             &0.312(2)    &0.017(4)         &12.3/11  \\
\\
                   20                 & -2                  & 31/44               &0.314(4)    &0.012(9)         &5.8/9   \\
\\
                   28                 & -2                  & 31/44               &0.314(6)    &0.013(14)         &4.7/7   \\
 \\ \hline\hline
\label{tab:C_parallel(L/2)}
\end{tabular}
\end{table*}

\begin{table*}[b]
\setlength{\tabcolsep}{8pt}
\caption{Fit to the finite-size perpendicular correlation function $C_\perp(L/2)$ data obtained from quantum Monte Carlo simulations at the special transition based on Eq.~\eqref{Eq:cor-Fss-perp-1}.}
\begin{tabular}{ccccccc}

\hline\hline
      $L_{min}$                       & $-2-\eta_\perp$    & $\hat{\eta}_{\perp_Q}$     & $a_1$    &$a_2$      &$\chi^2/d.o.f$ \\ \hline
\\
                   12                 & -2                 & 0.671(3)           & 0.338(2)   &0                &20.4/11    \\   
\\
                   20                 & -2                  & 0.679(6)           &0.334(3)     &0                &8.0/9    \\
\\
                   28                 & -2                  & 0.668(9)           &0.339(4)     &0                &5.0/7   \\
\\
                   12                 & -2.018(1)           & 8/11             &0.3355(10)    &0                &22.9/11  \\    
\\
                   20                 & -2.014(2)           & 8/11            &0.331(2)    &0                &7.1/9   \\
\\
                   28                 & -2.016(3)           & 8/11             &0.333(3)    &0                &4.4/7   \\
\\
                   12                 & -2                  & 8/11             &0.292(2)     &0.056(3)         &17.7/11 \\
\\
                   20                 & -2                  & 8/11              &0.294(3)    &0.051(6)         &9.2/9   \\
\\
                   28                 & -2                  & 8/11               &0.288(4)    &0.066(10)         &5.5/7   \\
 \\ \hline\hline
\label{tab:C_perp(L/2)}
\end{tabular}
\end{table*}

\end{document}